\documentstyle[12pt]{article}
\catcode `\@=11
\@addtoreset{equation}{section}

\def\qed{\vrule height 6pt width 6pt depth 6pt}

\newcommand{\be}{\begin{equation}}
\newcommand{\ee}{\end{equation}}

\newtheorem{thm}{Theorem}
\newtheorem{rem}{Remark}

\begin{document}

\begin{center}
{\bf \Large{ON A ORDER REDUCTION THEOREM\\
~\\IN THE LAGRANGIAN FORMALISM }}
\end{center}
\vskip 1.0truecm
\centerline{D. R. Grigore\footnote{e-mail:
grigore@theor1.ifa.ro, grigore@roifa.ifa.ro}}
\vskip 0.5 truecm
\centerline{ Dept. Theor. Phys., Inst. Atomic Phys.,}
\vskip 0.5 truecm
\centerline{Bucharest- M\u agurele, P. O. Box MG 6, ROM\^ANIA}
\bigskip \nopagebreak
\begin{abstract}
\noindent
We provide a new proof of a important theorem in the Lagrangian
formalism about necessary and sufficient conditions for a
second-order variational system of equations to follow from a
first-order Lagrangian.
\end{abstract}

\section{Introduction}

The importance of variational equations, i.e. equations which
can be obtained from a variational principle, is well
established in the physics literature. Practically all the
equations of physical interest (electromagnetism, Yang-Mills,
gravitation, string theory, etc.) are of this type. When faced
with a system of partial differential equations (the number of
the equations being equal to the number of the field
components), one has to answer two questions: 1) if the
equations are variational (at least locally); 2) how to choose
the ``simplest" Lagrangian i.e. of the lowest possible order. 

The first question is answered completely by the so-called
Anderson-Duchamp-Krupka \cite{AD}, \cite{K1} equations. These
equations give necessary and sufficient conditions on a set of
partial differential equations (the number of the equations being 
equal to the number of the field components) such that they can be 
derived from a (local) Lagrangian. These equations are nothing
else but the generalization to classical field theory and to
arbitrary order of the well-known Helmholtz-Sonin equations from
particle mechanics and for second-order equations. The proof is
based on the construction of an explicit Lagrangian associated
to any set of variational equations, called Tonti Lagrangian.
This Lagrangian is, in general, of the same order as the
equations, so it is highly degenerated. The second question
(called here {\it the order-reduction problem}) is not solved in
complete generality, although a number of partial results and
conjectures are available in the literature (see for instance
\cite{A} and references cited there). It is clear that if the
Lagrangian is of order
$r$
then the order of the equations will be 
$s = 2r$
or less (if the Lagrangian is degenerated). One can conjecture
that if the order of the equations is 
$s$
then one can choose a Lagrangian of minimal order
$
\left[ {s \over 2}\right] + 1.
$
This conjecture is true in particle mechanics \cite{OK} but, in
general, is false in classical field theory.

However, there exists a case which can be completely analyzed in
the classical field theory framework, namely the case of 
second-order equations. Needless to say, most of the physical
applications fall into this case. In \cite{AD} one can find the
proof of the following statement: a second-order system of
partial differential equations (the number of the equations being 
equal to the number of the field components) follows from a
first-order Lagrangian if and only if it is at most linear in
the second-order derivatives. This particular case is
nevertheless, extremely important for physical applications.
Indeed, in \cite{G1} and \cite{G2} one finds out the proof of
the following facts: (i) every locally variational second-order
system of partial derivative equations (the number of the equations 
being  equal to the number of the field components) is a
polynomial of maximal degree equal to the dimension of the space-time
manifold in the second-order derivatives. (In fact, the
dependence is through some special combinations called
hyper-Jacobians \cite{O}); (ii) a number of physically
interesting groups of Noetherian symmetries (like gauge
invariance or general transformation of coordinates in gravity)
have the effect of reducing the dependence of the second-order
derivatives to a dependence which is at most linear. So, in this
case we are exactly in the particular case of the
order-reduction theorem enounced above. 

The proof of this theorem from \cite{AD} is extremely long and
based on complicated computations. Because results of this type
are less known in the physical literature, we think that it is
interesting to provide a rather elementary proof of the
order-reduction statement from above. The idea of the proof is
to use complete induction. This type of argument was intensively
used in \cite{G1} and \cite{G2} and seems to be extremely useful
for further generalizations. We feel that this line of argument
might be the simplest in trying to analyze the general case.

In Section 2 we present the general formalism of jet-bundle
extensions applied to variational problems and in section 3 we
prove the order-reduction theorem. 

\section{Jet Bundle Extensions and Variational Calculus}

Suppose that
$
\pi: Y \mapsto X
$
is a fibration of the manifold 
$Y$
(of dimension
$N + n$)
over the manifold
$X$
(of dimension
$n$). 

If
$
x \in X, y \in Y
$
and
$
\zeta, \zeta': X \rightarrow Y
$
are smooth sections such that
$
\zeta(x) = \zeta'(x) = y
$
then we say that
$\zeta$
is equivalent to
$\zeta'$
if their partial derivatives up to order
$r$
computed in a arbitrary chart
$
(U,\phi)
$
around
$x$
and
$
(V,\psi)
$
around
$y$
are identical. Then we denote the equivalence class of
$\zeta$
by
$
j^{r}_{n}\zeta
$
and the set of all such equivalence classes by
$
J^{r}_{x.y}(Y).
$
Then the
$r$-order
jet bundle extension is by definition:

\be
J^{r}_{n}(Y) = \cup J^{r}_{x,y}(Y).
\ee

We denote by
$
\pi^{s,t}: J^{s}_{n}(Y) \rightarrow J^{t}_{n}(Y)~~~(t \leq s \leq r)
$
the canonical projections. By convention
$
J^{0}_{n}(Y) \equiv Y.
$

In the following we give the details for the case 
$
r = 2.
$
If
$
\phi = (x^{\mu}),~\mu = 1,...,n
$
and
$
\psi = (x^{\mu},\psi^{A}),~A = 1,...,N
$
are two chart systems adapted to the fiber bundle structure,
then one can extend it to
$
J^{2}_{n}(S): (V^{2},\psi^{2})
$
where
$
V^{2} \equiv (\pi^{2,0})^{-1}(V)
$
and
$
\psi^{2} = (x^{\mu},\psi^{A},\psi^{A}_{\mu},\psi^{A}_{\mu\nu})
$
where 
$
\mu \leq \nu.
$

The definition of the last two coordinates are:

\be
\psi^{A}_{\mu}(j^{2}_{n}\zeta) \equiv
\partial_{\mu}\psi^{A} \circ \zeta(x) \circ \phi^{-1} (\phi(x))
\ee

and 

\be
\psi^{A}_{\mu\nu}(j^{2}_{n}\zeta) \equiv
\partial_{\mu}\partial_{\nu} \psi^{A} \circ\zeta(x) \circ
\phi^{-1} (\phi(x)).
\ee

It is convenient to extend
$
\psi^{A}_{\mu\nu}
$
to all couples
$
\mu\nu
$
by symmetry: we denote

\be 
\{\mu\nu\} = \cases{ \mu\nu & for $\mu \leq \nu$ \cr
\nu\mu & for $\nu \leq \mu$ \cr}; 
\nonumber
\ee
then
$
\psi^{A}_{\mu\nu} \equiv \psi^{A}_{\{\mu\nu\}}.
$
Now we define the differential operators:

\be
\partial^{\mu}_{A} \equiv {\partial \over \partial \psi^{A}_{\mu}};~~~
\partial^{\mu\nu}_{A} \equiv {\partial \over \partial \psi^{A}_{\mu\nu}}
\times \cases{ 1 & if $\mu = \nu$ \cr 1/2 & if $\mu \not= \nu$ \cr}
\ee

and the total derivative operators

\be 
D_{\mu} \equiv {\partial \over \partial x^{\mu}} + \psi^{A}
\partial^{\mu}_{A} + \psi^{A}_{\nu} \partial^{\mu\nu}_{A} .
\ee

Suppose that
$
{\cal T}_{A},~(A = 1,...,N)
$
are some some smooth functions on
$
J^{2}_{n}(Y)
$
i.e. they depend on
$
(x^{\mu},\psi^{A},\psi^{A}_{\mu},\psi^{A}_{\mu\nu}).
$

One calls the 
$
n + 1
$-form

\be
T = {\cal T}_{A}~d\psi^{A} \wedge dx^{1} \wedge \cdots \wedge dx^{n}
\label{edif}
\ee
a second-order differential equation. One says that
$T$
is {\it locally variational} if there exists a locally defined function
${\cal L}$
on
$
J^{2}_{n}(Y)
$
such that:

\be
{\cal T}_{A} = {\cal E}_{A}({\cal L}) \equiv \left(\partial_{A} -
D_{\mu} \partial^{\mu}_{A} + D_{\mu}D_{\nu} \partial^{\mu\nu}_{A} \right)
{\cal L}
\label{Eop}
\ee

One calls
${\cal L}$
a {\it local Lagrangian} and:

\be
L \equiv {\cal L}~dx^{1}\wedge\cdots \wedge dx^{n}
\label{Lform}
\ee
a {\it local Lagrange form}.

If the differential equation
$T$
is constructed as above then we denote it by
$
E(L).
$
A local Lagrangian is called a {\it total divergence} if it is of the form:
\be
{\cal L} = D_{\mu} V^{\mu}.
\ee

One can check that in this case we have:
\be
E(L) = 0.
\label{trEL}
\ee

The content of the local variationality conditions ADK is
expressed by a set of partial derivative equations on the components
$
{\cal T}_{A}
$
which can be found in \cite{G1} (see eqs. (3.1)-(3.3) there). 
As we have said in the introduction, we will be concerned here
with the case when
$
{\cal T}_{A}
$
are of the form:

\be
{\cal T}_{A} = t^{\mu\nu}_{AB}~\psi^{B}_{\mu\nu} + t_{A}.
\label{EL}
\ee
with
$
t^{\mu\nu}_{AB}
$
and
$
t_{A}
$
smooth functions depending on
$
(x^{\mu},\psi^{A},\psi^{A}_{\nu})
$
and one can suppose that we have the symmetry property:

\be
t^{\mu\nu}_{AB} = t^{\nu\mu}_{AB}.
\label{eq1}
\ee

The ADK equations are in this case:

\be
t^{\mu\nu}_{AB} = t^{\mu\nu}_{BA}
\label{eq2}
\ee

\be
\partial^{\mu}_{A} t_{B} + \partial^{\mu}_{B} t_{A} =
2~{\delta \over \delta x^{\nu}} t^{\mu\nu}_{AB},
\label{eq3}
\ee

\be
\partial^{\rho}_{C} t^{\mu\nu}_{AB} + 
\partial^{\rho}_{B} t^{\mu\nu}_{AC} =
\partial^{\mu}_{A} t^{\rho\nu}_{BC} +
\partial^{\nu}_{A} t^{\rho\mu}_{BC},
\label{eq4}
\ee

\be
2 \partial_{B} t_{A} - {\delta \over \delta x^{\nu}}
\partial^{\nu}_{B} t_{A} = A \leftrightarrow B,
\label{eq5}
\ee

\be
4 \partial_{B} t^{\mu\nu}_{AC} - 2 {\delta \over \delta x^{\rho}}
\partial^{\rho}_{B} t^{\mu\nu}_{AC} -
(\partial^{\mu}_{B}  \partial^{\nu}_{C} +
\partial^{\nu}_{B}  \partial^{\mu}_{C}) t_{A} =
A \leftrightarrow B,
\label{eq6}
\ee

\be
(\partial^{\mu}_{B}  \partial^{\nu}_{C} +
\partial^{\nu}_{B}  \partial^{\mu}_{C})  t^{\rho\sigma}_{AD} =
(\partial^{\rho}_{A}  \partial^{\sigma}_{D} +
\partial^{\sigma}_{A}  \partial^{\rho}_{D}) t^{\mu\nu}_{BC}
\label{eq7}
\ee
where:

\be
{\delta \over \delta x^{\mu}} \equiv {\partial \over \partial
x^{\mu}} + \psi^{A} \partial^{\mu}_{A}.
\ee

If the equations (\ref{eq1})-(\ref{eq7}) are satisfied then the
differential equation
$T$
can be derived from a Lagrangian. This Lagrangian is highly
non-unique. A possible choise is the Tonti expression which is
in our case:
 
\be
{\cal L} = {\cal L}_{0} + {\cal L}_{1}
\label{Tonti1}
\ee
where

\be
{\cal L}_{0} \equiv \int_{0}^{1} \psi^{A} t_{A} \circ \chi_{\lambda}
d\lambda,
\label{Tonti2}
\ee
and

\be
{\cal L}_{1} \equiv \psi^{A}_{\mu\nu} {\cal L}^{\mu\nu}_{A}
\label{Tonti3}
\ee
with

\be
{\cal L}^{\mu\nu}_{A} = \int_{0}^{1} \lambda \psi^{B} t^{\mu\nu}_{AB}
\circ \chi_{\lambda} d\lambda
\label{Tonti4}
\ee
and

\be
\chi_{\lambda} (x^{\mu}, \psi^{A}, \psi^{A}_{\nu}) =
(x^{\mu},\lambda \psi^{A}, \lambda\psi^{A}_{\nu}).
\label{Tonti5}
\ee

Notice that
${\cal L}$
is a second-order Lagrangian. The purpose of this paper is to
prove that one can find a equivalent Lagrangian which is of
first order.

Let us note in closing that from (\ref{eq4}) and (\ref{eq7}) one
obtains that the functions
$
{\cal L}^{\mu\nu}_{A}
$
defined above satisfy the following equations:

\be
\partial^{\rho}_{B} {\cal L}^{\mu\nu}_{A} +
\partial^{\mu}_{B} {\cal L}^{\rho\nu}_{A} +
\partial^{\nu}_{B} {\cal L}^{\mu\rho}_{A} = A \leftrightarrow B
\label{int1}
\ee
and

\be
\partial^{\mu}_{B} (\partial^{\rho}_{C} {\cal L}^{\sigma\nu}_{A} +
\partial^{\sigma}_{C} {\cal L}^{\sigma\rho}_{A}) +
\partial^{\nu}_{B} (\partial^{\rho}_{C} {\cal L}^{\sigma\mu}_{A} +
\partial^{\sigma}_{C} {\cal L}^{\sigma\mu}_{A}) -
(\partial^{\mu}_{A} \partial^{\nu}_{B} +
\partial^{\nu}_{A} \partial^{\mu}_{B}) {\cal L}^{\sigma\rho}_{C} -
(\partial^{\rho}_{A} \partial^{\sigma}_{C} +
\partial^{\sigma}_{A} \partial^{\rho}_{C}) {\cal L}^{\mu\nu}_{B}
= 0.
\label{int2}
\ee

\begin{rem}

In \cite{AD} these two equations are obtained in a different
way: one computes
$
{\cal E}_{A}({\cal L})
$
with
$
{\cal L}
$
given by (\ref{Tonti1})-(\ref{Tonti5}) and imposes a structure
of the type (\ref{EL}).
\end{rem}

\section{The Order Reduction Theorem}

In this section we will prove by induction the following theorem:

\begin{thm}

$
(T_{n})
$
~~~Suppose that the functions
$
{\cal L}^{\mu\nu}_{A}
$
depending on
$
(x^{\mu},\psi^{A},\psi^{A}_{\nu})
$
verify the equations (\ref{int1}) and (\ref{int2}). Then there
exists a set of functions
$
V^{\mu}
$
depending on the same variables such that:

\be
{\cal L}^{\mu\nu}_{A} = \partial^{\mu}_{A} V^{\nu} + 
\partial^{\nu}_{A} V^{\mu}.
\label{AD}
\ee
\end{thm}

Before starting the proof of theorem
$
(T_{n})
$
let us note the following corollary:

\begin{thm}

$
(C_{n})
$
~~~If the functions
$
t^{\mu\nu}_{AB}
$
verify the system of equations (\ref{eq1}), (\ref{eq2}),
(\ref{eq4}) and (\ref{eq7}) then there exists a function
$
{\cal L}
$
depending on
$
(x^{\mu},\psi^{A},\psi^{A}_{\nu})
$
such that:

\be
t^{\mu\nu}_{AB} = -{1 \over 2} 
\left(\partial^{\mu}_{A} \partial^{\nu}_{B} +
\partial^{\nu}_{A} \partial^{\mu}_{B} \right) {\cal L}.
\label{int3}
\ee
\end{thm}

{\bf Proof of the corollary:}

Define
$
{\cal L}_{1}
$
as in (\ref{Tonti3})-(\ref{Tonti5}). Because (\ref{eq4}) and
(\ref{eq7}) are valid we also have (\ref{int1}) and
(\ref{int2}). Applying
$
(T_{n})
$
we obtain the functions
$
V^{\mu}
$
such that (\ref{AD}) is true. Next, we define:

\be
{\cal L}' \equiv {\cal L}_{1} - 2 D_{\mu} V^{\mu}
\ee
and prove that
$
{\cal L}'
$
does not depend on the second-order derivatives
$
\psi^{A}_{\mu\nu}
$
i.e. is a first-order Lagrangian. Moreover, because
$
D_{\mu} V^{\mu}
$
is a trivial Lagrangian we have:

\be
{\cal E}_{A}({\cal L}_{1}) = {\cal E}_{A}({\cal L}').
\nonumber 
\ee

So we have:

\be
{\cal T}_{A} = {\cal E}_{A}({\cal L}) = {\cal E}_{A}({\cal L}_{0}) 
+ {\cal E}_{A}({\cal L}_{1}) = {\cal E}_{A}({\cal L}_{0}) +
{\cal E}_{A}({\cal L}') = {\cal E}_{A}({\cal L}_{0}+{\cal L}')
\nonumber
\ee
so we have (\ref{int3}) with
$
{\cal L} \rightarrow {\cal L}_{0} + {\cal L}'
$
which is of first order.
$\qed$

We start now the induction proof of
$
(T_{n}).
$
The assertion
$
(T_{1})
$
is elementary: from (\ref{int1}) we have in this case:

\be
\partial_{B} {\cal L}_{A} = \partial_{A} {\cal L}_{B}
\nonumber
\ee
and using Frobenius theorem one obtains a (local) function
$V$,
depending on the variables
$
(x^{\mu},\psi^{A},\psi^{A}_{\nu})
$
such that:

\be
{\cal L}_{A} = 2 \partial_{A} V
\nonumber
\ee
i.e. we have  (\ref{AD}) for 
$
n = 1.
$

Suppose that
$
(T_{n})
$
is true: we prove
$
(T_{n+1}).
$
It is convenient to assume that the 
$
n + 1
$
indices run from
$0$
to
$n;$
then the indices from
$1$
to
$n$
are denoted by latin letters
$
i, j,...
$
and the indices from
$0$
to
$n$
by greek letters
$
\alpha, \beta,...
$

We divide the proof in a number of steps.

(i) We take in (\ref{int1})
$
\mu = \nu = \rho = 0
$
and we have:

\be
\partial_{B} {\cal L}^{00}_{A} = \partial_{A} {\cal L}^{00}_{B}
\label{ind1}
\ee
and Frobenius theorem provides us with a function
$
V^{0}
$
of
$
(x^{\mu},\psi^{A},\psi^{A}_{\nu})
$
such that:

\be
\partial^{0}_{A} V^{0} = {1 \over 2} {\cal L}^{00}_{A} 
\label{V01}
\ee
i.e. we have (\ref{AD}) for
$
\mu = \nu = 0.
$
One can give an explicit expression for
$
V^{0}
$
using a well-known homotopy formula:

\be
V^{0} = {1\over 2} \int_{0}^{1} \psi^{C}_{0} {\cal L}^{00}_{C}
\circ \phi_{s} ds + \tilde{V}^{0}.
\label{V0}
\ee

Here

\be
\phi_{s}(x^{\mu},\psi^{A},\psi^{A}_{\nu}) = 
(x^{\mu},\psi^{A},s\psi^{A}_{0},\psi^{A}_{i})
\ee
and
$
\tilde{V}^{0}
$
is a function independent of 
$
\psi^{A}_{0}
$
i.e.

\be
\partial^{0}_{A} \tilde{V}^{0} = 0.
\label{C1}
\ee

For the moment
$
\tilde{V}^{0}
$
is restricted only by this relation.

(ii) Let us define now the functions:

\be
T_{AB}^{ij} \equiv \partial^{i}_{B} {\cal L}^{0j}_{A} +
\partial^{j}_{B} {\cal L}^{0i}_{A} - \partial^{0}_{A} {\cal L}^{ij}_{B}.
\label{T}
\ee 

We will prove that
$
\tilde{V}^{0}
$
can be chosen such that
$
V^{0}
$
fulfills, beside (\ref{V01}) the equation:

\be
(\partial^{i}_{A} \partial^{j}_{B} + \partial^{j}_{A} \partial^{i}_{B})
V^{0} = T_{AB}^{ij}.
\label{V02}
\ee

Indeed, if we introduce here 
$
V^{0}
$
given by (\ref{V0}) we obtain that
$
\tilde{V}^{0}
$
verifies an equation of the type:

\be
(\partial^{i}_{A} \partial^{j}_{B} + \partial^{j}_{A} \partial^{i}_{B})
\tilde{V}^{0} = \tilde{T}_{AB}^{ij}
\label{V01a}
\ee
with

\be
\tilde{T}_{AB}^{ij} \equiv T_{AB}^{ij} - {1\over 2} \int_{0}^{1}
\psi^{C}_{0} \left[ 
(\partial^{i}_{A} \partial^{j}_{B} + \partial^{j}_{A} \partial^{i}_{B})
{\cal L}^{00}_{C}\right] \circ \phi_{s} ds.
\ee

The equation (\ref{V01a}) involves in the left-hand side a function
$
\tilde{V}^{0}
$
which does not depend on 
$
\psi^{A}_{0}
$
(see (\ref{C1})). The same must be true for the right-hand side.
Indeed, using (\ref{ind1}) and (\ref{int2}) for
$
\mu = i,~\nu = j,~\rho = \sigma = 0
$
one easily proves that:

\be
\partial^{0}_{D} \tilde{T}_{AB}^{ij} = 0.
\ee

So, the system (\ref{V01a}) makes sense. We now note that this
system is exactly of the same type as in the corollary
$
(C_{n}).
$
So, one has to check if the integrability conditions of the type
(\ref{eq1}),(\ref{eq2}), (\ref{eq4}) and (\ref{eq7}) hold and
one is entitled to apply the induction procedure. Indeed we have:

- (\ref{eq1}): directly from the definition (\ref{T})

- (\ref{eq2}): from (\ref{int1}) with
$
\mu = i,\nu = j,\rho = 0
$

- (\ref{eq4}): by direct computation using (\ref{int1}) and (\ref{int2})

- (\ref{eq7}): again by direct computation using (\ref{int2}).

So, applying the induction hypothesis and the implication
$
(T_{n}) \Rightarrow (C_{n})
$
one gets that the system (\ref{V01a}) has a solution. We will
suppose from now on that
$
V^{0}
$
is fixed (non-uniquely however) by (\ref{V01}) and (\ref{V02}).

(iii) We now define the functions:

\be
G^{i}_{A} \equiv {\cal L}_{A}^{0i} - \partial^{i}_{A} V^{0}.
\label{G}
\ee

One proves using (\ref{int2}) that:

\be
\partial^{0}_{B} G^{i}_{A} = \partial^{0}_{A} G^{i}_{B} .
\ee

Using Frobenius theorem one obtains the existence of a set of
(local) functions
$
V^{i}
$
such that:

\be
\partial^{0}_{A} V^{i} = G^{i}_{A}.
\label{Vi}
\ee

Explicitly, one has a formula of the same type as (\ref{V0}):

\be
V^{i} = \int_{0}^{i} \psi^{A}_{0} G^{i}_{A} \circ \psi_{s} +
\tilde{V}^{i} 
\label{Vi1}
\ee
where:

\be
\partial^{0}_{A} \tilde{V}^{i} = 0.
\ee

From (\ref{G}) and (\ref{Vi}) we note that it follows:

\be
\partial^{0}_{A} V^{i} + \partial^{i}_{A} V^{0} = {\cal L}^{0i}_{A}
\ee
i.e. (\ref{AD}) for
$
\mu = i,~\nu = 0.
$

(iv) Let us define the functions:

\be
\tilde{\cal L}^{ij}_{A} \equiv {\cal L}^{ij}_{A} - \int_{0}^{1}
\psi^{C}_{0} (\partial^{i}_{A} G^{j}_{C} + \partial^{j}_{A}
G^{i}_{C}) \circ \phi_{s} ds.
\label{L}
\ee

One can check by direct computation that:

\be
\partial^{0}_{A}  \tilde{\cal L}^{ij}_{A} = 0.
\ee

Moreover, it follows from (\ref{int1}) and (\ref{int2}) that
$
\tilde{\cal L}^{ij}_{A}
$
verifies integrability equations of the same type as
$
{\cal L}^{ij}_{A}.
$
So, applying the induction procedure we obtain a set of functions
$
\tilde{V}^{i}
$
depending on
$
(x^{\mu},\psi^{A},\psi^{A}_{i})
$
such that:

\be
\tilde{\cal L}^{ij}_{A} = \partial^{i}_{A} \tilde{V}^{j} +
\partial^{j}_{A} \tilde{V}^{i}.
\ee

It is natural to take in (\ref{Vi1})
$
\tilde{V}^{i}
$
exactly the solution of the system above. Then (\ref{L}) implies:

\be
{\cal L}^{ij}_{A} = \partial^{i}_{A} V^{j} + \partial^{j}_{A} V^{i}
\ee
i.e. we have (\ref{AD}) for
$
\mu = i,~\nu = j.
$
This finishes the induction.
$\qed$

Finally we note that the same idea used in proving
$
(T_{n}) \Rightarrow (C_{n})
$
proves the order-reduction theorem i.e. if
$
{\cal T}_{A}
$
given by (\ref{EL}) is locally variational, then one can find a
first-order Lagrangian
${\cal L}$
such that
$
{\cal T}_{A} = {\cal E}_{A}({\cal L}).
$

In particular we have (\ref{int3}) and:

\be
t_{A} = \partial_{A} {\cal L} - {\delta \over \delta x^{\mu}}
\partial^{\mu}_{A} {\cal L}.
\ee

\begin{rem}

The function
${\cal L}$
is determined by (\ref{int3}) up to an expression of the
following form:

\be
{\cal L} = \sum_{k=0}^{n} {1\over k!}
C^{\mu_{1},...,\mu_{k}}_{A_{1},...,A_{k}}
\prod_{I=0}^{k} \psi^{A_{i}}_{\mu_{i}}
\ee
where the functions
$
C^{\mu_{1},...,\mu_{k}}_{A_{1},...,A_{k}}
$
are independent of the first order-derivatives and are completely
antisymmetric in the indices
$
\mu_{1},...,\mu_{k}
$
and in the indices
$
A_{1},...,A_{k}.
$

This is the first step in deriving the most general expression
of a trivial first-order Lagrangian (see \cite{K2}).
\end{rem}

\end{document}